\documentclass{rendiconti_davide}

\usepackage{amsmath,amssymb}
\usepackage{mathpazo} 
\usepackage{bm}
\usepackage{graphicx,subfigure,color}
\usepackage{fancyhdr}
\usepackage{pdfpages}
\usepackage{verbatim}
\definecolor{linkcolor}{rgb}{0.0,0.3,0.5}
\usepackage[unicode,colorlinks=true, linkcolor=linkcolor, citecolor=linkcolor, filecolor=linkcolor,urlcolor=linkcolor]{hyperref}
\usepackage[all]{hypcap} 
\usepackage[all]{nowidow}
\usepackage{titlesec}
\titleformat{\subsection}
  {\large\bfseries}{\thesubsection}{1em}{}
\usepackage{titlesec}
\titleformat{\section}
  {\Large\bfseries}{\thesection}{1em}{}

\usepackage[authoryear]{natbib}

\fancypagestyle{plain}{
\fancyhf{}
\fancyfoot[R]
\thepage

}
\newcommand\orcid[1]{\href{https://orcid.org/#1}{$\!\!$\includegraphics[scale=0.006]{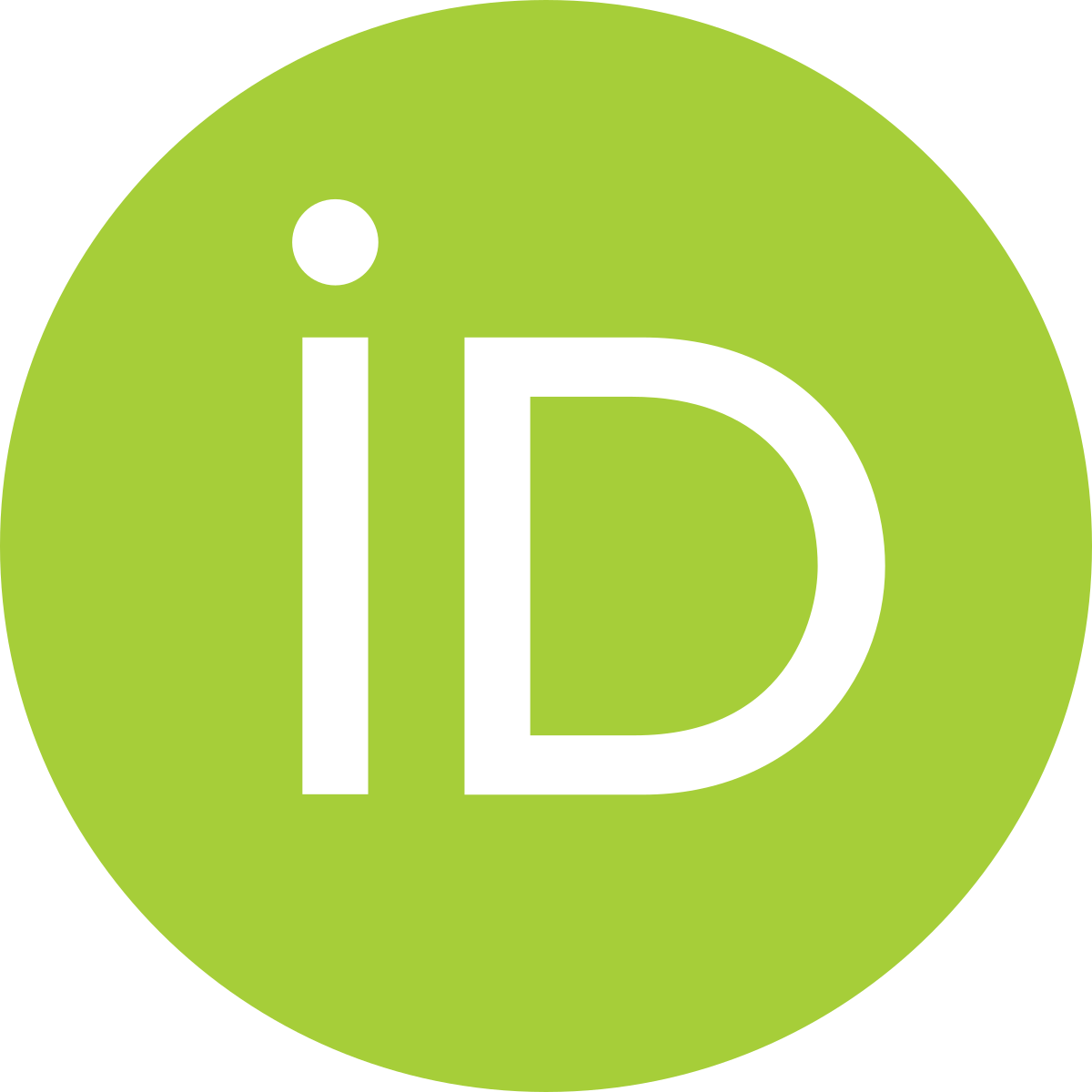} $\!\!$}}
\definecolor{Mygrey}{gray}{0.75}
\newcommand{\ssim}{\mathchar"5218\relax\,}

\title{Modeling the outcome of supernova explosions in binary population synthesis using the stellar compactness}
\author{{\bf{
Maciej Dabrowny$\;\;$\orcid{0000-0003-0391-4919}$\,^{1,*}$, 
Nicola Giacobbo$\;\;$\orcid{0000-0002-8339-0889}$\,^{1}$,
Davide Gerosa$\;\;$\orcid{0000-0002-0933-3579}$\,^{1\dagger}$
}}
\vspace{1cm}
\\
{\bf Abstract}
\vspace{0.1cm}\\
{\normalsize
Following the collapse of their cores, some of the massive binary stars that populate our Universe are expected to form merging binaries composed of black holes and neutron stars. Gravitational-wave observations of the resulting compact binaries can reveal precious details on the inner workings of the supernova mechanism and the subsequent formation of compact objects. Within the framework of the population-synthesis code {\sc MOBSE}, we present the implementation of a new supernova model that relies on the  compactness of the collapsing star. The model has two free parameters, namely the compactness threshold that separates the formation of black holes and that of neutron stars, and the fraction of the envelope that falls back onto the newly formed black holes. We compare this model extensively against other prescriptions that are commonly used in binary population synthesis. 
We find that the cleanest signatures of the role of the pre-supernova stellar compactness are (i) the relative formation rates of the different kinds of compact binaries, which mainly depend on the compactness threshold parameter, and (ii) the location of the upper edge of the mass gap between the lightest black holes and the heaviest neutron stars, which mainly depends on the fallback fraction.
 \vspace{0.3cm}
\vspace{0.5cm}
}
}

\begin{document}

\maketitle	

\let\thefootnote\relax\footnote{
\begin{affiliations}\rm \small
\vspace{-0.2cm}

\item School of Physics and Astronomy \& Institute for Gravitational Wave Astronomy, University of Birmingham, Birmingham, B15 2TT, UK.

 $*$ \href{mailto:mxd773@star.sr.bham.ac.uk}{mxd773@star.sr.bham.ac.uk}

\end{affiliations}
}

\vspace{-1cm}

\section{Introduction} \label{sec:intro}

Gravitational wave (GW) observations of merging compact binaries offer unprecedented insights into the life of massive stars. The black holes (BHs) and neutron stars (NSs) observed by LIGO and Virgo \citep{2019PhRvX...9c1040A,2021PhRvX..11b1053A} constitute the end product of stellar collapse ---the same cosmic events that are well understood to be behind supernova (SN) explosions. The direct observation of compact binaries thus provides a novel opportunity to probe the inner workings of the SN explosion mechanism. 

Stars with zero-age-main-sequence (ZAMS) masses greater than $\ssim 10$ M$_\odot$ are robustly predicted to undergo core-collapse SN once their cores reach the Chandrasekhar limit (for a recent review on SN theory see \citealt{2021Natur.589...29B}). Lighter stars might instead explode as electron-capture SN. In those cases, the degenerate oxygen-neon core reaches the critical mass of $\ssim 1.38$ M$_\odot$ and electron-capture reactions destabilize the inner region %
\citep{1980PASJ...32..303M,1984ApJ...277..791N}. While NSs are the only possible outcome of electron-capture SNe, core-collapse SNe can also produce BHs, with the mass of the resulting compact remnant increasing with the ZAMS mass overall. For progenitors heavier than $\ssim 70$ $M_\odot$, however, the collapsing core becomes unstable to pair production. This removes radiation pressure support from the star which, in turn, ignites explosive carbon-oxygen (CO) burning. The core is either partially (pulsational pair-instability SN;  \citealt{2007Natur.450..390W}) or entirely (pair-instability SN; \citealt{2003ApJ...591..288H}) disrupted, thus introducing a characteristic upper limit of $\ssim 50 M_\odot$  \citep{2019ApJ...887...53F,2021ApJ...912L..31W} to the masses of BH remnants that can be produced by conventional stellar evolution.

The key GW observables to probe the physics of SN are the so-called 
\emph{gaps} in the BH mass spectrum. The existence or absence of compact objects with masses between $\ssim 3 M_\odot$ and $\ssim 5 M_\odot$ (``lower mass gap'') separating BHs and NSs has been long investigated using X-ray binary data~\citep{1998ApJ...499..367B,2010ApJ...725.1918O,2011ApJ...741..103F} and is now under active scrutiny after some of the most recent LIGO/Virgo events, notably GW190814 \citep{2021PhRvX..11b1053A}. At the high-end of the BH mass spectrum (``upper mass gap''), current GW measurements point to a significant drop in the merger rate which is consistent with prediction from pair-instability SN theory. At the same time, some of the observed BH masses, notably those of GW190521 \citep{2021PhRvX..11b1053A}, are well within this forbidden region, perhaps hinting at a different origin for this event (for a review see \citealt{2021NatAs.tmp..136G}). Besides the masses, the orientations of the BH spins can also provide insights on SN physics. In particular, they are sensitive to the asymmetric emission of mass and neutrinos occurring during the SN and the subsequent kick imparted to the newly formed compact object (e.g. \citealt{2000ApJ...541..319K}; \citealt{2013PhRvD..87j4028G,2017PhRvL.119a1101O}).

In this paper, we consider the standard formation scenario where GW sources originate from massive binary stars (e.g. \citealt{2014LRR....17....3P}) and explore the impact of a new SN model on the resulting compact binaries. The most common prescriptions used to predict the SN outcome implemented in state-of-the-art population-synthesis codes are based on the CO mass after the carbon burning stage and the pre-supernova total mass of the progenitor star %
(e.g. {\sc SeBa}, \citealt{1996A&A...309..179P}; {\sc StarTrack}, \citealt{2020A&A...636A.104B}; {\sc COSMIC}, \citealt{2020ApJ...898...71B}; {\sc MOBSE}, \citealt{2018MNRAS.474.2959G}; {\sc COMPAS}, \citealt{2017NatCo...814906S}).

In particular, the two leading SN models are the so-called {\it rapid} and {\it delayed} prescriptions first proposed by \cite{2012ApJ...749...91F}. The main difference between these two models is the time after core bounce at which the explosion is launched, which is set to $< 250$ ms and $> 500$ ms in the rapid and delayed case, respectively. In terms of the properties of the resulting BH binaries, the rapid (delayed) model does (does not) predict the lower mass gap between BHs and NSs (e.g. \citealt{2012ApJ...757...91B}). Additional recipes to model the SN outcome include the effect of envelope stripping in binaries \citep{2021A&A...645A...5S} and probabilistic couplings between natal kicks and remnant masses \citep{2020MNRAS.499.3214M}.

Hydrodynamical simulations (e.g. \citealt{2011ApJ...730...70O}) suggest %
that, for a given equation of state, the most critical parameter to estimate the outcome of a SN is given by the compactness of the stellar core at the onset of the explosion,  
\begin{equation}
\label{eq:compactness}
    \zeta_{\rm M} = \frac{M/M_\odot}{R(M)/1000\,{\rm km}},
\end{equation}
where $R(M)$ is the radius that encloses a mass $M$ at core bounce. 

\begin{figure*}[t]
    \centering
    \includegraphics{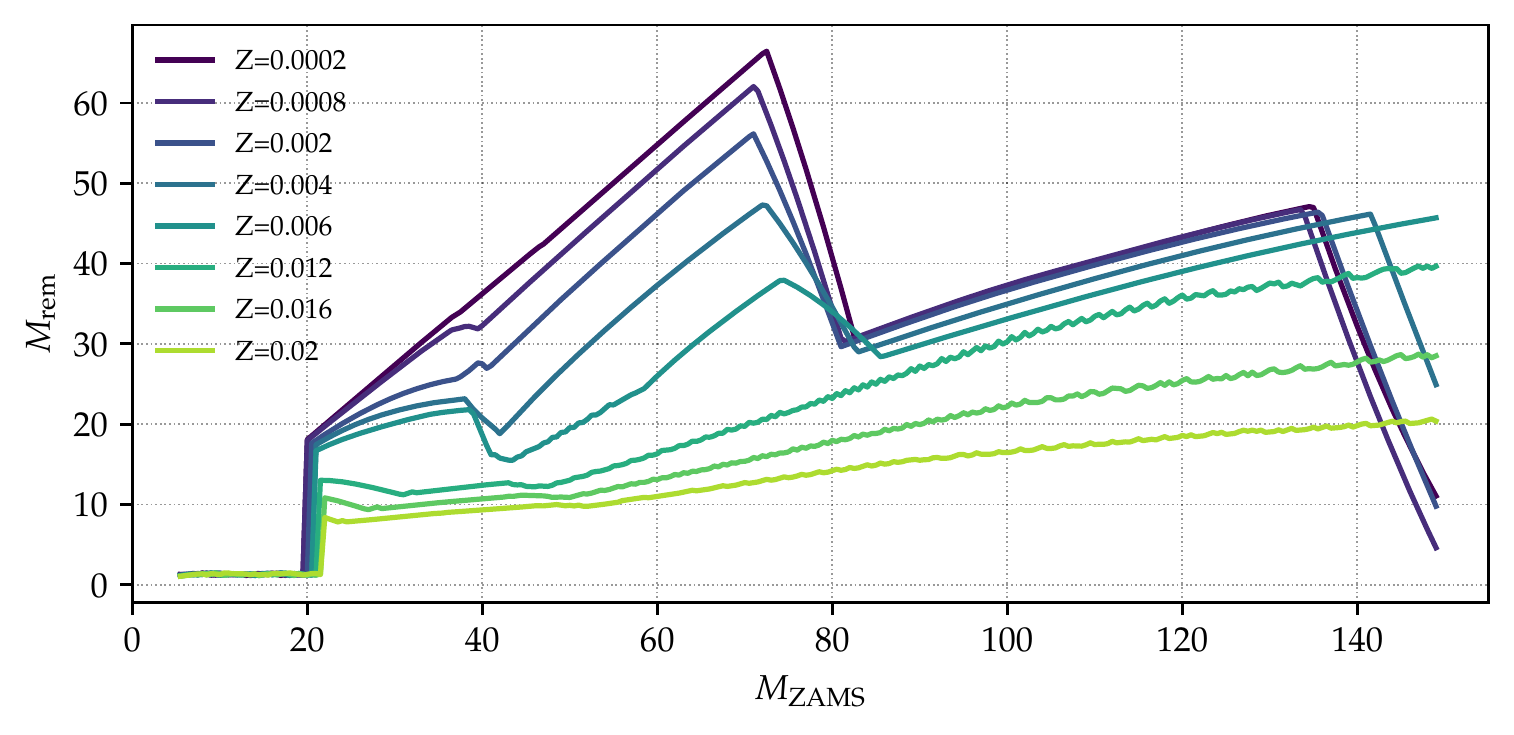}
    \caption{Mass spectrum of compact remnants as a function of the progenitor initial mass $M_{\rm ZAMS}$ in our fiducial compactness SN model ($\zeta_{2.5}=0.365$ and $f_{\rm H}=0.9$). Different colors represent different metallicities $Z=0.0002, \dots, 0.02$ (dark to light).}
    \label{fig:spec}
\end{figure*}

We explore the impact of such ``compactness model'' of compact-object formation on the resulting single and binary mass spectrum by means of population-synthesis simulations. In Sec.~\ref{sec:methods}, we describe an updated version of the {\sc MOBSE} code where we have implemented a treatment of SNe based on the stellar compactness. %
In Sec.~\ref{sec:results}, we discuss the effect of the compactness model on the formation and evolution of merging compact binaries and compare our results against the more common rapid and delayed prescriptions. In Sec.~\ref{sec:conclusions}, we summarize our findings and illustrate prospects for future work.

\section{Methods} \label{sec:methods}

After a brief introduction to the  {\sc MOBSE} code (Sec.~\ref{mobseintro}), we describe the implementation of our compactness model (Sec.~\ref{sec:method2}) and the simulation setup used to explore its relevance (Sec.~\ref{sec:methods.sims}). 

\subsection{Massive Objects in Binary Stellar Evolution}
\label{mobseintro}
Both single and binary stars are evolved with the rapid population-synthesis code {\sc MOBSE}
(``Massive Objects in Binary Stellar Evolution'', \citealt{2018MNRAS.474.2959G}). Built on top of the older {\sc BSE} \citep{2002MNRAS.329..897H}, {\sc MOBSE}'s key updates include up-to-date prescriptions for stellar winds in massive stars \citep{2018MNRAS.480.2011G}, natal kicks \citep{2019MNRAS.482.2234G,2020ApJ...891..141G}, and modeling of pulsational pair-instability and pair-instability SNe \citep{2017MNRAS.470.4739S}. 
The version of {\sc MOBSE}  presented in this paper has been assigned number v1.1. %
The code is publicly available at \href{https://mobse-webpage.netlify.app/about/}{mobse-webpage.netlify.app}.

\subsection{Compactness model}
\label{sec:method2}
While the prescriptions for the rapid and delayed models are straightforward to implement in a population-synthesis code, the compactness parameter $\zeta_{M}$ of Eq.~(\ref{eq:compactness})  depends on the internal structure of the star before the SN. This information is typically not available with a population-synthesis approach, which does  not track the details of stellar structure but only some of the star's global properties. 

As in \cite{2011ApJ...730...70O}, we will refer to the reference value of $\zeta$ at $M=2.5 M_\odot$, resulting in the parameter $\zeta_{\rm 2.5}$.  \cite{2018ApJS..237...13L} demonstrated that there is a strong correlation between $\zeta_{2.5}$ and the carbon-oxygen mass $M_{\rm CO}$ of the pre-SN, which is a readily accessible quantity. More recently, \cite{2020ApJ...888...76M} showed that this relation can be well represented by the simple fit
\begin{equation}
\label{eq:zeta}
    \zeta_{2.5} \simeq 0.55 - 1.1 \left( \frac{M_{\rm CO}}{1\, M_{\odot}}\right)^{-1}\,.
\end{equation}

We make use of $\zeta_{2.5}$ to distinguish between the formation of a NS and that of a BH. The precise threshold value is still debated, with current estimates ranging from $\ssim 0.2$ \citep{2014MNRAS.445L..99H} to  $\ssim 0.45$ \citep{2011ApJ...730...70O}. In our fiducial model, we assume a treshold value of $\zeta_{2.5} = 0.365$ which was chosen to match the NS-to-BH transition predicted by the rapid model (cf. Sec.~\ref{sec:single}). Hence, stellar progenitors with $\zeta_{2.5}  > 0.365$ will form BHs %
while progenitors with $\zeta_{2.5}  \leq 0.365$ will form NSs. 

The compactness itself, however, does not predict the mass of the resulting compact remnant. We thus adopt the same strategy of \cite{2020ApJ...888...76M}. NS masses are drawn from a Gaussian distribution with mean $\mu = 1.33$ M$_\odot$ and standard deviation $\sigma = 0.09$ M$_\odot$, which agrees with the observed masses of NSs in binary systems \citep{2016ARA&A..54..401O}. BH masses are instead given by
\begin{equation}
\label{eq:fh}
    M_{\rm BH} = M_{\rm He} + f_{\rm H}(M_{\rm fin} - M_{\rm He}),
\end{equation}
where $M_{\rm He}$ is the mass of the helium core and $M_{\rm fin}$ is the total stellar mass at the onset of collapse (both of which are tracked in standard population synthesis). The quantity $f_{\rm H}
\in[0,1]$  is a free parameter that describes the fraction of the hydrogen envelope that is accreted by the BH. We adopt a fiducial value $f_{\rm H}=0.9$, i.e. assuming that %
 90\% of the star's hydrogen envelope falls back onto the remnant after core collapse. This fiducial value is motivated by recent studies \cite[e.g.][]{2013ApJ...769..109L,2016ApJ...821...38S,2018MNRAS.476.2366F} %
 that show that not all of the hydrogen envelope is accreted onto the newborn BH, even in the case of direct collapse. This is because the outermost layers of the envelope are only weakly bound to the core. %

\begin{figure}[t]
    \centering
    \includegraphics[scale=0.8]{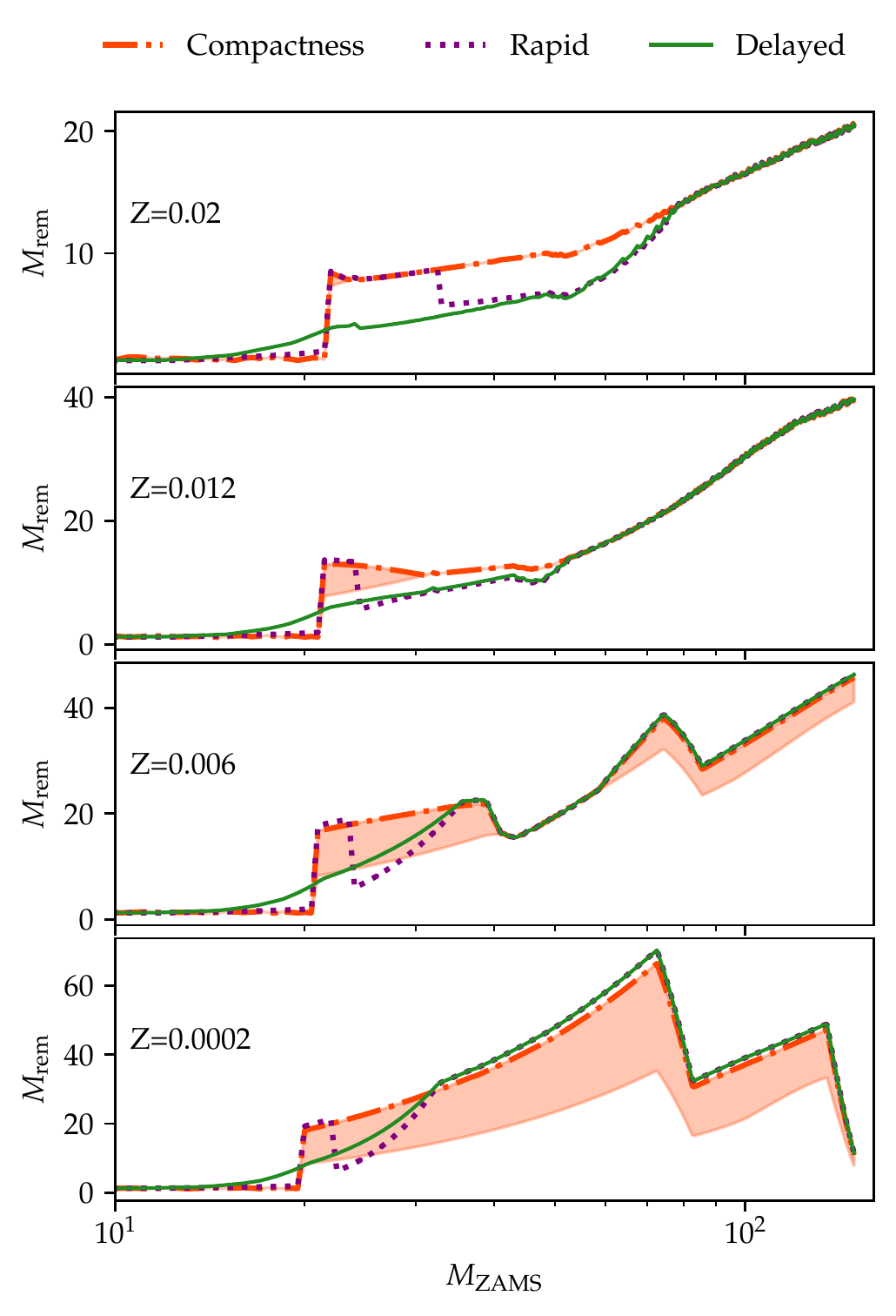}
    \caption{Comparison between different SN explosion models. Red dash-dotted, purple dotted and green solid lines mark the fiducial compactness model, the rapid model, and the delayed model, respectively. Each panel illustrates how the mass of the remnant $M_{\rm rem}$ depends on the mass of the initial star $M_{\rm ZAMS}$ for different metallicities $Z$. The shaded areas indicate the range of masses obtained with $f_{\rm H}=0.9$ (fiducial, thick red lines) and $f_{\rm H}=0.1$ (thin orange lines). In this figure, the NS/BH compactness threshold is fixed to $\zeta_{2.5}=0.365$.}
    \label{fig:modelComparison}
\end{figure}

Figure~\ref{fig:spec} illustrates the relationship between the initial mass of the star $M_{\rm ZAMS}$ and the mass of the compact remnant $M_{\rm rem}$ obtained with our fiducial compactness model ($\zeta_{2.5}  = 0.365$ and $f_{\rm H}=0.9$). Much like in the rapid case, the compactness model also produces a gap in the remnant masses  which separates BHs and NSs.  As already explored at length \citep{2010ApJ...714.1217B,2015MNRAS.451.4086S,2018MNRAS.474.2959G,2019MNRAS.490.3740N}, both the maximum BH mass and the upper edge of the mass gap strongly depends on the progenitor's metallicity $Z$. This is because higher metallicities drive larger mass loss via stellar winds, thus leading to smaller remnant masses. The peak at $M_{\rm ZAMS}\sim 70 M_\odot$ for $Z\lesssim 0.004$ is due to pulsational pair-instability SNe while the sharp decrease at  $M_{\rm ZAMS}\sim 140 M_\odot$ marks the onset of the proper pair-instability regime.%

\subsection{Simulation setup}
\label{sec:methods.sims}

 Starting from the fiducial model we just described, 
we further explore the parameter space spanned by $\zeta_{2.5}$ and $f_{\rm H}$. We consider four different values for the threshold parameter $\zeta_{2.5} =$ 0.2, 0.3, 0.365 (fiducial), and 0.4. For each of these, we vary $f_{\rm H} =$ 0.1, 0.3, 0.5, 0.7, and 0.9 (fiducial). For all the resulting combinations of $\zeta_{2.5}$ and $f_{\rm H}$ we evolve 10$^6$ massive binaries varying their metallicity $Z=$ 0.0002, 0.0004, 0.0008, 0.0012, 0.0016, 0.002, 0.004, 0.006, 0.008, 0.012, 0.016, and 0.02. We also run control cases using  the rapid and delayed models. The details of their precise implementation in {\sc MOBSE} have been presented by \cite{2021MNRAS.502.4877S} and include some minor modifications with the respect to the original prescriptions by \cite{2012ApJ...749...91F}. %

The initial condition of the simulated binaries are generated as follows \citep{2018MNRAS.480.2011G}: %
primary ZAMS masses are extracted from an initial mass function (IMF) $p(M_1)\propto M_1^{-2.3}$ with $M_1 \in [5,150] M_\odot$,   mass ratios $q=M_2/M_1$ are drawn according to $p(q|M_1) \propto q^{-0.1}$ with $q \in [0.1,1]$, periods $P$ are drawn according to $p(P) \propto \log(P/{\rm days})^{-0.55}$ with $P\in [10^{0.15},10^{5.5}]$, and eccentricities are drawn according to $p(e) \propto e^{-0.45}$ with $e \in [0,1]$ %
\citep{2012Sci...337..444S}. %

\begin{figure}[t]
    \centering
    \includegraphics[scale=0.8]{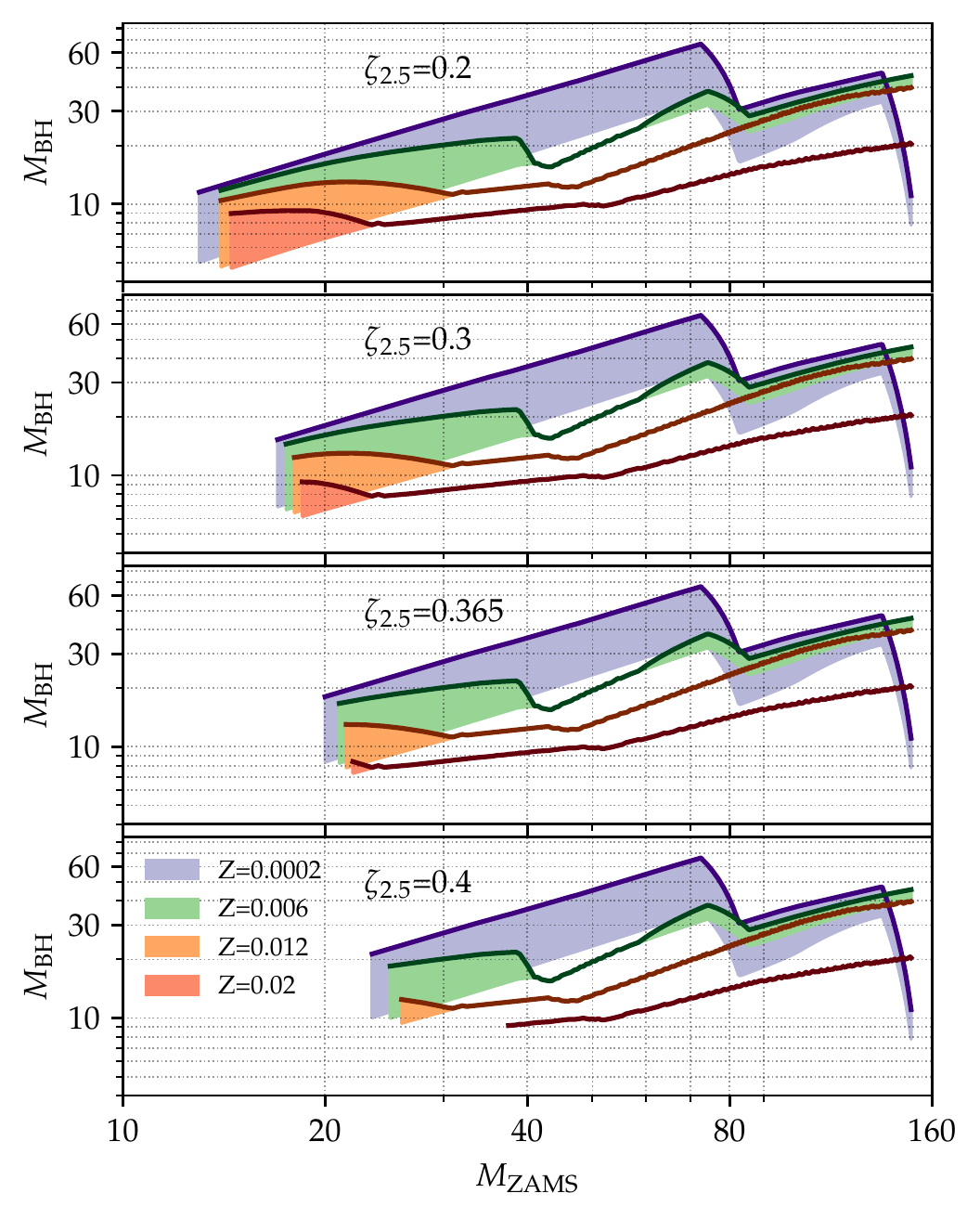}
    \caption{BH mass spectrum as a function of the initial mass $M_{\rm ZAMS}$ for different metallicities $Z=$ 0.02, 0.012, 0.006,  and 0.0002 (colors). Each panel is produced assuming a different value of the NS/BH compactness threshold $\zeta_{2.5}=$ 0.2, 0.3, 0.365, 0.4 (top to bottom). The shaded areas indicate the range of masses obtained with $f_{\rm H} = 0.9$ (fiducial, thick lines) and $f_{\rm H} = 0.1$ (lower edge of the colored regions). 
    }
    \label{fig:zetafh}
\end{figure}

\section{Results}\label{sec:results}
We now present the results of our simulations. We first discuss the impact of the compactness SN model on the mass spectrum of compact object (Sec.~\ref{sec:single}). We then illustrate the formation of compact object binaries (Sec.~\ref{sec:COB}) and  focus on the sub-sample of merging systems (Sec.~\ref{sec:mergers}). %

\begin{figure*}[t]
  \centering
  \includegraphics[width=1.\linewidth]{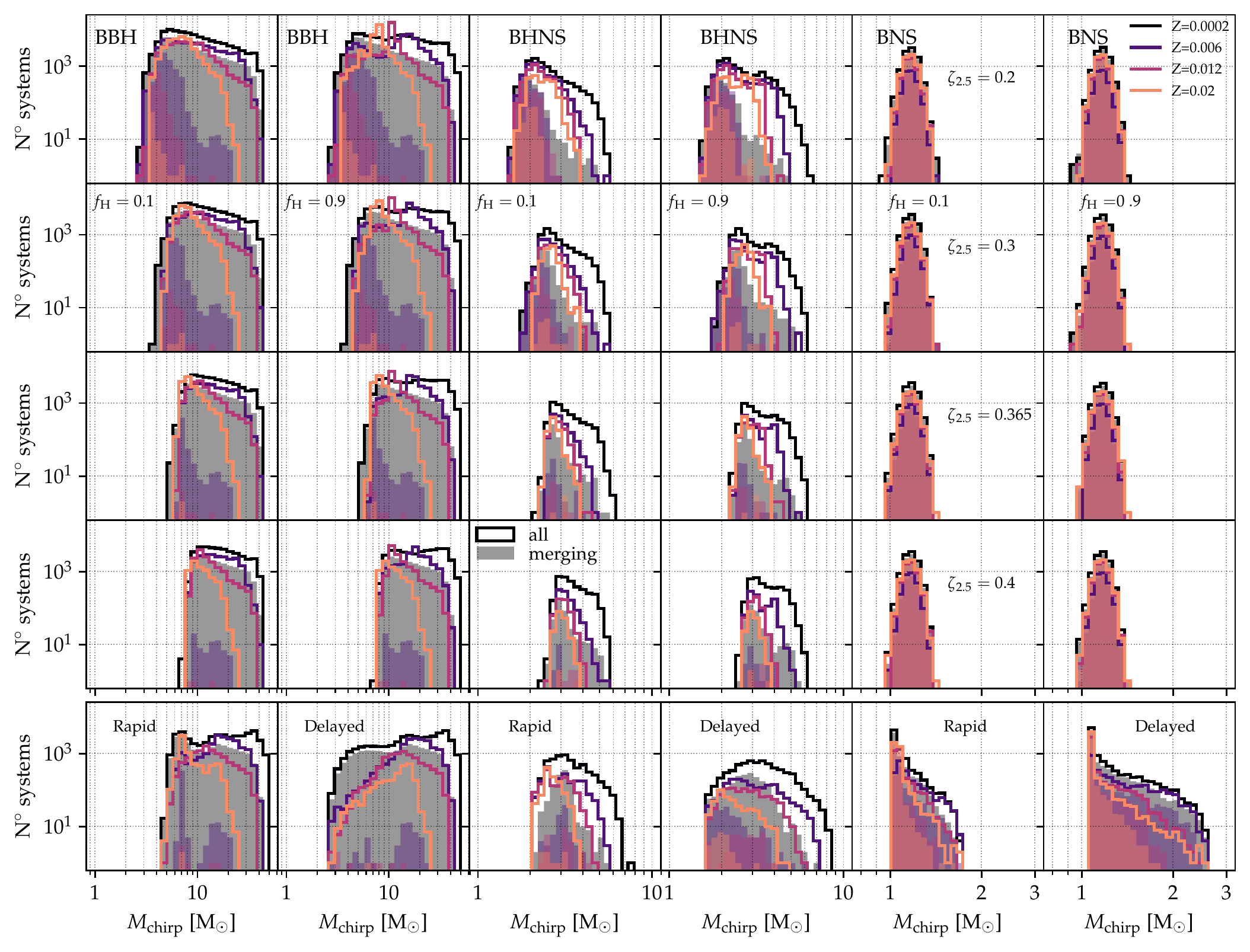}
\caption{Distribution of chirp masses $M_{\rm chirp}$ of BBHs (two leftmost columns), BHNSs (two middle columns), and BNSs (two rightmost columns). The top four rows show results obtained with the compactness model introduced in this paper, with $\zeta_{2.5}=$ 0.2, 0.3, 0.365, 0.4 increasing from top to bottom. The left (right) column of each pair is obtained with $f_{\rm H}=0.1$ ($f_{\rm H}=0.9$).  The bottom row shows control results from the standard rapid and delayed SN models. In all cases, colors indicate the metallicity, with $Z=$ 0.0002, 0.006, 0.012, 0.02 from darker to lighter. We further separate between the full sample of  compact binaries produced in our evolutions (empty histograms) and the subset of those that merge within a Hubble time (filled histograms).
}

\label{fig:cob}
\end{figure*}

\subsection{Mass spectrum}\label{sec:single}
In Fig.~\ref{fig:modelComparison}, we compare the mass spectrum for our fiducial compactness model ($\zeta_{2.5} = 0.365$, $f_{\rm H}=0.9$) against those obtained with the more standard rapid and delayed models. Both the rapid and the compactness models produce a gap in the remnant masses. However, while in the rapid model the mass gap  ranges from $\sim 2 M_\odot$ to $\sim 5 M_\odot$ independently of the metallicity, in the compactness model the size of the mass gap strongly depends on both $Z$ and $f_{\rm H}$. In particular, the shaded areas in Fig.~\ref{fig:modelComparison} indicate the range of predicted masses by compactness models with $0.1\leq f_{\rm H} \leq 0.9$. The impact of $f_{\rm H}$ is largest at lower metallicities. This is because stars at higher metallicities tend to end their life as Wolf-Rayet stars that have light hydrogen envelopes, so the mass of the resulting BH is almost independent of $f_{\rm H}$. On the other hand, at lower metallicities stellar winds are less efficient and stars approach the SN phase with heavier %
envelopes.

Furthermore, the compactness model does not predict a sharp decrease in $M_{\rm rem}$ at $20 M_\odot \gtrsim M_{\rm ZAMS} \gtrsim 30 M_\odot$ which instead characterizes the rapid model. %
With some dependency on metallicity and fallback, the compactness model tends to, overall, predict heavier remnants in that intermediate region (while for the rapid model those stars do not undergo direct collapse, \citealt{2012ApJ...749...91F}). At the high-end of the mass spectrum $M_{\rm zams}\gtrsim 50 M_\odot$ (but the precise value depends on $Z$), the rapid, delayed and fiducial compactness model all predict the formation of BHs via direct collapse and thus return similar remnant masses.

The four panels of Fig.~\ref{fig:zetafh} illustrate the impact of  $\zeta_{2.5}$ on the mass spectrum. As $\zeta_{2.5}$ is reduced, the NS-to-BH transition moves toward lower ZAMS masses, i.e. BHs are formed from lighter progenitors. This has important consequences for the relative abundances of NSs and BHs (cf. Sec.~\ref{sec:COB}). This is also sensitive to the metallicity. At $Z=0.02$ ($Z=0.0002$) the ZAMS mass transition value %
changes  from $\ssim 14 M_\odot $ ($\ssim 13 M_\odot$) for $\zeta_{2.5}=0.2$ to $\ssim 37 M_\odot$ ($\ssim 23 M_\odot$) for $\zeta_{2.5}=0.4$. 

Finally, let us notice that $f_{\rm H}$ has a strong impact on the maximum BH mass. Smaller values of $f_{\rm H}$ lead to lighter BHs by construction, because the parameter $f_{\rm H}$ describes the amount of fallback material.  More interestingly, for massive progenitors ($\gtrsim 60$ M$_\odot$) and lower values of $f_{\rm H}\sim 0.1$ %
, the dependence of the maximum BH mass on the metallicity is not monotonic. In particular, the heaviest BHs are formed at intermediate metallicities $Z \sim 0.006$.  

\subsection{Compact binaries}\label{sec:COB}
From our samples of simulated massive stars (see Sec.~\ref{sec:methods.sims}), we select the systems that form binaries composed of compact objects and classify them as binary black holes (BBHs), black-hole neutron-star  binaries (BHNSs), and binary neutron stars (BNSs). The solid lines in Fig.~\ref{fig:cob} shows the resulting distribution of their chirp masses $M_{\rm chirp} = {(M_1M_2)^{3/5}}/{(M_1 + M_2)^{1/5}}$ (where here $M_{1,2}$ are the masses of the two compact objects). This is the mass combination parameter that is measured best in GW observations (e.g. \citealt{1987thyg.book..330T}). Larger values of $\zeta_{2.5}$ result in a narrower mass range for both BBHs and BHNSs. For BBHs, the minimum chirp mass ranges from $\ssim 2.5$ M$_\odot$ for $\zeta_{2.5}=0.2$ to $\ssim 7$ M$_\odot$ for $\zeta_{2.5}=0.4$. In particular, for $\zeta_{2.5} = 0.365$ ($\zeta_{2.5} = 0.2$) our compactness model matches the range predicted by the rapid (delayed) prescription. The observationally based NS mass prescription we implemented  (cf. Sec.~\ref{sec:method2}) does not depend on $\zeta_{2.5}$. Consequently, the effect of $\zeta_{2.5}$ on the BHNS chirp  masses is mitigated  compared to BBHs, while the masses of BNSs are entirely unaffected. 

Figure~\ref{fig:cob} shows results for our two most extreme cases of fallback retention, $f_{\rm H}=$ 0.1 and 0.9, for each type of compact binaries and values of $\zeta_{2.5}$. As already highlighted in Sec.~\ref{sec:method2}, the impact of $f_{\rm H}$ on the remnant masses depends on both $Z$ and $\zeta_{2.5}$.
In particular, the mass spectrum of lighter stars ($M_{\rm ZAMS} \lesssim 40$) presents a shallower slope for higher $f_{\rm H}$. This translates ito well defined peaks in the chirp masses of BBHs for $f_{\rm H}=0.9$.
At low metallicities, the chirp-mass distribution of both BBHs and BHNSs shows a shortage of the most massive systems for $f_{\rm H}=0.1$ compared to $f_{\rm H}=0.9$. This is a direct consequence of the fact that, for larger ZAMS masses, the parameter $f_{\rm H}$ only affects BH masses at low metallicities.

\begin{figure}[t]
    \centering
    \includegraphics[scale=0.8]{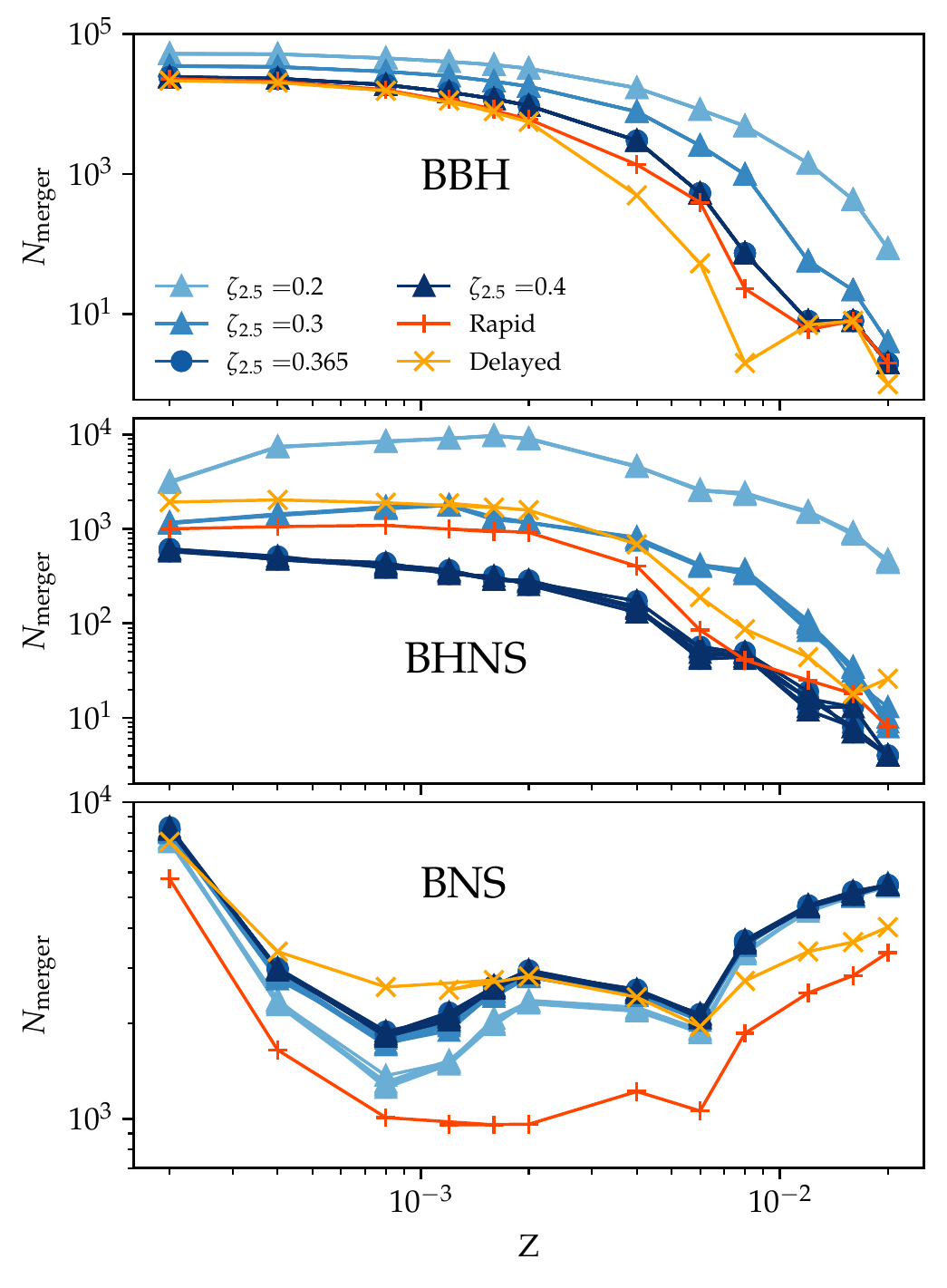}
    \caption{Number of compact binary mergers as a function of metallicity for BBHs (top panel), BHNSs (middle panel), and BNSs (bottom panel).  Results from the compactness model are shown in different shades of blue for $\zeta_{2.5}=0.2,0.3, 0.365,0.4$ (light to dark). The fiducial model with $\zeta_{2.5}=0.365$ is marked with circles, while the additional model variations are marked with triangles. Red pluses and yellow crosses indicate the standard rapid and delayed models, respectively.  For all the compactness models, results obtained for $f_{\rm H}=0.1, 0.3, 0.5, 0.7,$ and 0.9 are largely indistinguishable on the scale of this plot.
    }
    \label{fig:nmerger}
\end{figure}

\subsection{Merging systems}\label{sec:mergers}
Only a fraction of the compact binaries that form will merge within a Hubble time (here taken to be $13.6$ Gyr). In Fig.~\ref{fig:cob}, we compare the chirp masses of such merging systems against those of the entire simulated samples. In agreement with \cite{2018MNRAS.480.2011G,2019MNRAS.485..889S,2019ApJ...885....1W}%
, we find that the most massive systems do not merge within a Hubble time and that the fraction of merging systems decreases for increasing metallicities.

Figure~\ref{fig:nmerger} further illustrates the total number of merging systems as a function of metallicity, separating the different kinds of compact binaries (BBHs, BHNSs, and BNSs). The total number of merging BBHs and merging BHNSs strongly depends on the metallicity of the progenitors. In particular, most models show a monotonic trend that predicts more mergers at lower metallicities. The BHNSs formed in models with $\zeta_{2.5} \lesssim 0.3$ are an exception and tend to have a mild peak at intermediate metallicities. %
In contrast, the metallicity has a weaker impact on the BNSs  rate, with fewer mergers predicted at intermediate metallicities $Z\sim 10^{-3}$.

The number of merging BNSs tends to slightly decrease as $\zeta_{2.5}$ increases, while that of both BBHs and BHNSs increases substantially, especially at high metallicities. This happens because the NS-to-BH transition moves toward higher masses for higher values of $\zeta_{2.5}$ and this effect is more prominent for higher metallicities (cf. Fig.~\ref{fig:zetafh}). Because the IMF is bottom-heavy, the location of the NS-to-BH transition has a strong impact on the relative formation rates when outcomes are classified in terms of BBHs, BHNSs, and BNSs.

The fraction of merging system is largely independent of the fallback parameter $f_{\rm H}$. Indeed, results from simulations with different values of $f_{\rm H}$ overlap almost perfectly in Fig.~\ref{fig:nmerger} (at least within the resolution of the figure). 
In order to merge within a Hubble time, most systems need to evolve through episode(s) of mass transfer and/or common envelope, with a consequent loss of their hydrogen envelope. The progenitors of merging binaries are most likely Wolf-Rayet stars with light envelopes and, consequently, the resulting BHs are relatively unaffected by $f_{\rm H}$.

Finally, it is interesting to note that compactness model predicts a larger fraction of merging BBHs compared to either the rapid or the delayed models for all values of $\zeta_{2.5}$ and $f_{\rm H}$. 
This is related to the IMF, which in our case is a power law with negative spectral index. Moving the NS-to-BH transition at smaller ZAMS masses favors the formation of a larger number of BHs, which in turn translates into a larger number of (merging) BBHs. The merger time due to GW emission also depends on the component masses, with massive systems merging in a shorter time~\citep{1964PhRv..136.1224P}. 
This effect is also present in the case of merging BNSs: the rapid model tends to produce lighter NSs and thus has the lowest number of merging BNSs for all metallicities. This also explains why all runs performed with the compactness model have similar distributions of merging BNSs.
We find that the effect of $\zeta_{2.5}$ is strongest for BHNSs. This because shifting the formation threshold between NSs and BHs toward higher masses affects the kicks imparted to the newly formed compact objects. In models with larger $\zeta_{2.5}$,  relatively massive stars that would have formed BHs now collapse to NSs, with a consequent large mass ejection during the explosion. Conservation of linear momentum then translates the large mass of the ejecta into strong natal kicks with a consequent suppression of the merger rate.

\section{Conclusions} \label{sec:conclusions}
SNe are a key process to understand the formation of compact binaries, with different explosion mechanics  producing different observable features (e.g. the presence of mass gaps). In this work, we have investigated the impact of the pre-SN stellar compactness \citep{2011ApJ...730...70O} on the mass spectrum of compact objects and the resulting population of GW sources. In particular, we presented a new version of the {\sc MOBSE}  population-synthesis code that includes a new model of BH and NS formation. Our ``compactness  model'' has two free parameters \citep{2020ApJ...888...76M}:

\begin{itemize}
\item
The quantity $f_{\rm H}$ describes the fraction of the hydrogen envelope that falls back onto BHs after their formation. BHs are lighter for smaller $f_{\rm H}$ while, in our simple model at least, the masses of NSs are unaffected. This might have important consequences when combining GW sources formed from isolated binaries to those formed in dynamical formation channels (e.g. \citealt{2019ApJ...886...25B,2021ApJ...910..152Z,2021PhRvD.103h3021W}). For compactness models with small $f_{\rm H}$, the heaviest BHs in the sample become exclusive to dynamical channels where, for example, more massive objects can be assembled via repeated mergers \citep{2021NatAs.tmp..136G}.
\item
The parameter $\zeta_{2.5}$ instead marks the stellar compactness where stars transition from forming NSs to forming BHs. We find that this parameter mostly affects the relative abundance of BHs and NSs that are expected to form in realistic populations of massive stars. 
\end{itemize}

We compared our models against two of the most commonly used prescriptions for SN explosions in binary population synthesis: the rapid and delayed models first introduced by \cite{2012ApJ...749...91F}. In particular, we find that distinguishing between these models and our compactness model with GW data will be facilitated by the detection of BHs which are relatively light ($M\lesssim 30 M_\odot$). For heavier BHs, predictions from most of the models we tested tend to overlap. The cleanest signature of the compactness model is the location of the upper edge of the lower mass gap between the lightest black holes and the heaviest neutron stars, which depends %
mainly on $f_{\rm H}$, together with the metallicity $Z$. The merger rates might also  help us discriminate between the different models, although this claim needs to be verified with an analysis which includes the metallicity-dependent star-formation history and the LIGO/Virgo selection effects.  

The compactness model we implemented in {\sc MOBSE} suffers from several simplifications. First, we use the stellar compactness only to discriminate between NSs and BHs, but then rely on ad-hoc prescriptions for their masses. Furthermore, we use a monotonic expression for the compactness as a function of the CO mass, see Eq.~(\ref{eq:zeta}), 
although recent studies have shown that their interplay is more complex \citep{2018ApJ...860...93S,2020ApJ...890...43C,2021ApJ...916...79C}. Both these aspects will be refined in future work together with estimates of the resulting merger and detection rates. %
The consequences of the SN models presented in this paper  in terms of GW emission and heavy-element nucleosynthesis also need to be further explored.

\subsection*{Acknowledgements} 

We thank Chamini Gnanavel for discussions.
M.D. acknowledges support of a Physics of Future Days Meeting Grant of the American Physical Society.
D.G. and N.G. are supported by European Union's H2020 ERC Starting Grant No. 945155---GWmining, Leverhulme Trust Grant No. RPG-2019-350, and Royal Society Grant No. RGS-R2-202004.  
D.G. is honored for having received the 2020 Minister of Cultural Heritage research prize of the Lincei Italian National Academy.
Computational work was performed on the University of Birmingham BlueBEAR cluster. 

\bibliographystyle{rendiconti_davide}
\setlength{\bibsep}{1pt}
{
\section*{References}\vspace{-0.7cm}
{\rm \bibliography{compactness}}
}

\end{document}